\documentclass[conference]{IEEEtran}

\providecommand{\keywords}[1]{\textbf{\textit{Keywords---}} #1}

\usepackage{amssymb}
\usepackage{graphicx}
\usepackage{multirow}
\usepackage{multicol}
\usepackage{caption}
\usepackage{fancyvrb}
\usepackage{url}
\usepackage{cite}
\usepackage[utf8]{inputenc}
\usepackage{subfig}

\usepackage{siunitx}
\makeatletter
\def\ps@IEEEtitlepagestyle{%
  \def\@oddfoot{\mycopyrightnotice}%
  \def\@evenfoot{}%
}
\def\mycopyrightnotice{%
  {\footnotesize 978-1-5386-2524-8/17/\$31.00 \textcopyright 2017 IEEE\hfill}
  \gdef\mycopyrightnotice{}
}

\begin{document}

\title{\LARGE \bf Characterizing Geo-located Tweets in Brazilian Megacities}

\author{
	\IEEEauthorblockN{
		João Pereira\IEEEauthorrefmark{1}\IEEEauthorrefmark{2},
		Arian Pasquali\IEEEauthorrefmark{3},
		Pedro Saleiro\IEEEauthorrefmark{1}\IEEEauthorrefmark{2},
		Rosaldo Rossetti\IEEEauthorrefmark{1}\IEEEauthorrefmark{2} and
		Nélio Cacho\IEEEauthorrefmark{4}}
	\IEEEauthorblockA{
		Artificial Intelligence and Computer Science Lab\IEEEauthorrefmark{1},\\
		Faculty of Engineering, University of Porto\IEEEauthorrefmark{2},\\
		INESC TEC\IEEEauthorrefmark{3}, Porto, Portugal\\
		Universidade Federal do Rio Grande do Norte\IEEEauthorrefmark{4}, Rio Grande do Norte, Brasil\\
		Email: joao.filipe.pereira@fe.up.pt, arrp@inesctec.pt, pssc@fe.up.pt, rossetti@fe.up.pt and neliocacho@dimap.ufrn.br}}

\maketitle

\begin{abstract}
This work presents a framework for collecting, processing and mining geo-located tweets in order to extract meaningful and actionable knowledge in the context of smart cities. We collected and characterized more than 9M tweets from the two biggest cities in Brazil, Rio de Janeiro and S\~ao Paulo. We performed topic modeling using the Latent Dirichlet Allocation model to produce an unsupervised distribution of semantic topics over the stream of geo-located tweets as well as a distribution of words over those topics. We manually labeled and aggregated similar topics obtaining a total of 29 different topics across both cities. Results showed similarities in the majority of topics for both cities, reflecting similar interests and concerns among the population of Rio de Janeiro and S\~ao Paulo. Nevertheless, some specific topics are more predominant in one of the cities.
\end{abstract}


\IEEEpeerreviewmaketitle

\section{Introduction}
With the rise of Social Media, people obtain and share information almost instantly on a 24/7 basis. Many research areas have tried to extract valuable insights from these large volumes of user generated content. The research areas of intelligent transportation systems and smart cities are no exception. Transforming this data into valuable information can be meaningful and useful for city governance, transportation services or even ordinary citizens wanting to be constantly informed about their cities. 

Among the existing social networks, Twitter is probably the most adequate for these purposes due to its microblog nature in which users publicly share short messages about their daily life. Other social networks, such as Facebook are not so accessible as users tend to publish content within a private circle of friends. Twitter is the 11th most visited website\footnote{\url{http://www.alexa.com/topsites}} in the planet. Its community is continuously growing and, nowadays, the number of active users is about 313 million\footnote{\url{https://about.twitter.com/company}}, registering a daily average of 400 million new posts. Nevertheless, extracting meaningful and actionable knowledge from Twitter is a complex endeavor. First,  the volume of tweets produced can be overwhelming for automatic collection, processing and mining, and second, tweets are usually short, informal, with a lot of abbreviations, jargon and idioms.

Taking the aforementioned challenges in consideration, we developed a framework for collecting, processing and mining of geo-located tweets with the goal of extracting knowledge from social media streams that might be useful in the context of smart cities. Although only 1-2\% of all tweets published every day are geo-located ~\cite{ikeda2013twitter}, these tweets have the advantage of having an explicit geographic relevance to the city where users published those messages. The framework provides a statistical analysis of the geo-located tweets regarding volume and content frequency over time. Content analysis is performed using Latent Dirichlet Allocation~\cite{blei2003latent} to extract the distribution of semantic topics from the geo-located Twitter stream. Furthermore, we support manual annotation of those topics in order to obtain meaningful descriptions and being able to compare topics across different cities. 

In this work, we use our framework to monitor geo-located tweets in two Brazilian megacities, Rio de Janeiro (RJ) and São Paulo (SP), with 6.5M and 12M inhabitants, respectively. Brazil is an ideal scenario to perform this kind of analysis. It has the two biggest cities in South America and Twitter is widely popular in Brazil with more than 17M active users. Moreover, both Rio de Janeiro and  São Paulo are in the top 10 cities with the higher percentage of geo-located tweets~\cite{leetaru2013mapping}. We collected more than 9M geo-located tweets in both cities for a period of 2 months, from March 2017 to May 2017. We identified 29 unique semantic topics and created a temporal distribution to compare both cities. To the best of our knowledge, it is the first study of these dimension comparing geo-located tweets of Rio de Janeiro and São Paulo.

The remainder of the paper is organized as follows. We present related work on content analysis of social media and the data collection process. We perform an exploratory analysis of the data collected for both cities and then we describe the topic modelling approach. We end with the results and analysis followed by conclusions and future developments.

\section{Related work}\label{related_work}

The majority of works focused on social media data, in particular, Twitter data, try to explore the content of the messages. Recent researches explored geotagged tweets in areas that range from sentiment analysis~\cite{musto2015developing,ulloa2016mining} to trend tracking~\cite{kwak2010twitter}. Musto et al.~\cite{musto2015developing} applied a domain-agnostic framework to two different scenarios, one related to a catastrophe and other related to the intolerance level in Italy. The framework is capable of collect data in a specific area and analyses the sentiment present in the content. In the work by Ulloa et al.~\cite{ulloa2016mining}, authors also proposed a framework capable of extracting meaningful pieces of information from real-time geo-located Twitter data in the field of transportation.

One of the first studies made using Twitter data was proposed by Kwak et al.~\cite{kwak2010twitter} and consisted of the collection of messages to classify the trends in its content. Results showed that almost 80\% of the trends on Twitter are related to real-time news and the period in which each trend maintains itself at the top is limited. Within this, Twitter can be seen as a mirror of real-time occurring events/incidents in the world.

Several works were already proposed to identify social patterns in the population daily-basis life and mapping such patterns geographically by topic modelling techniques to discover latent topics in social media streams. Usually, studies about topic modelling, in particular, LDA model, to text mining problems follow unsupervised approaches~\cite{lansley2016geography,oliveira2016sentibubbles} - where is not required the creation of a training dataset. Others improved the model and made it an supervised approach~\cite{ramage2010characterizing}, dependent on training data, and compare to the traditional one in order to prove better results.

Using entity-centric aggregations and topic modelling techniques, Oliveira et al.~\cite{oliveira2016sentibubbles} built a system focused on data visualization that allows a user to search for an entity during a specific period and shows which are the main topics identified in the Twitter messages.  Ordinary weekday patterns were identified by Lansley et al.~\cite{lansley2016geography} in their study regarding the inner region of London. The authors used an LDA model to distribute 20 topics over 1.3M tweets. After crossing the results of the experiment with land-uses datasets it was possible to observe interesting patterns in specific zones and places of the British city. Nonetheless, Ramage et al.~\cite{ramage2010characterizing} improved an LDA model by adding a supervised layer that automatic label each tweet used in their experiment.

\section{Data}
In this section, it is described the chosen approach to collect the data for this study. Such approach is based on the Twitter Streaming API, particularly, in the Python open-source library Tweepy\footnote{\url{http://www.tweepy.org/}}, which was configured enabling the 'location' filter. The main purpose of this type of configuration is the retrieving of all tweets within a defined bounding-box, i.e. a rectangle composed by two pair of latitude and longitude coordinates (South-West and North-East).

\subsection{Collecting Geo-located Tweets}

\begin{figure}[!ht]
\centering
\includegraphics[width=0.65\linewidth]{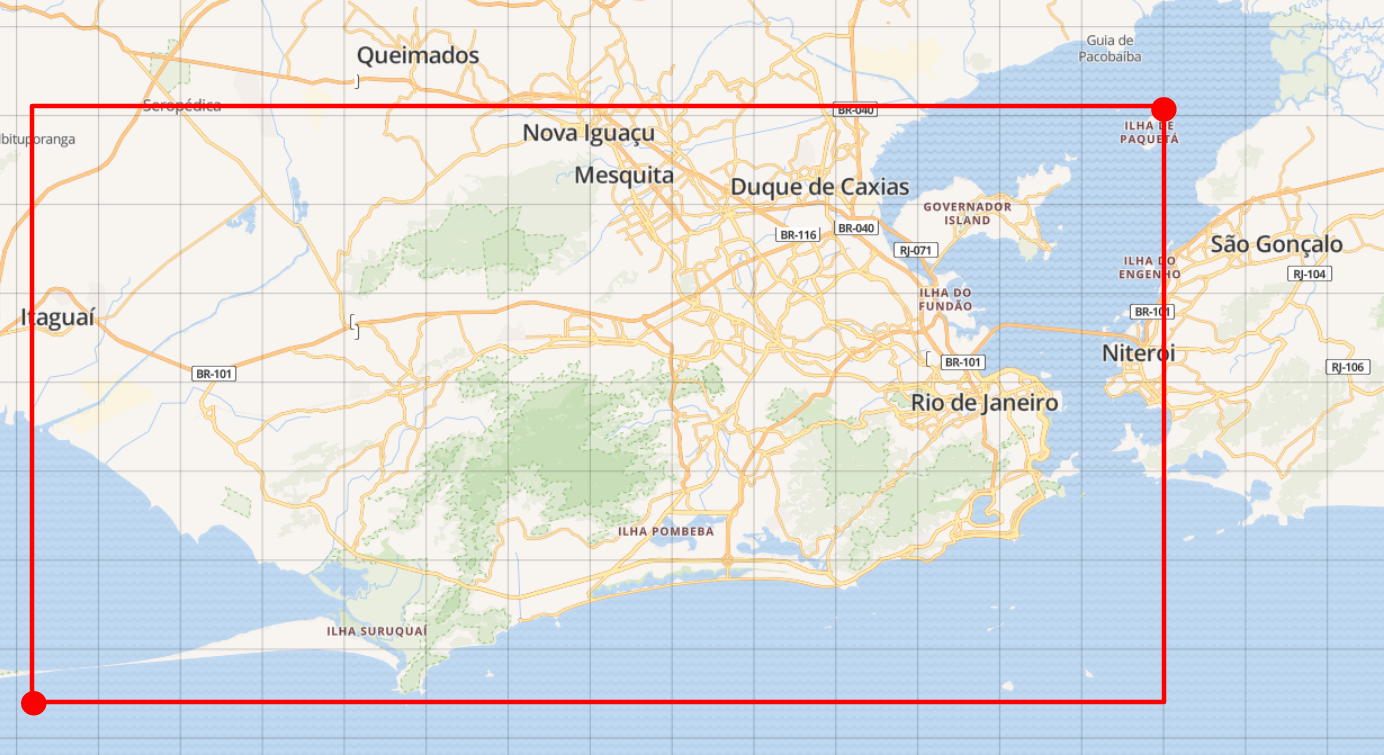}
\caption{Rio de Janeiro}
\label{fig:bb_rio}

\hfill

\includegraphics[width=0.65\linewidth]{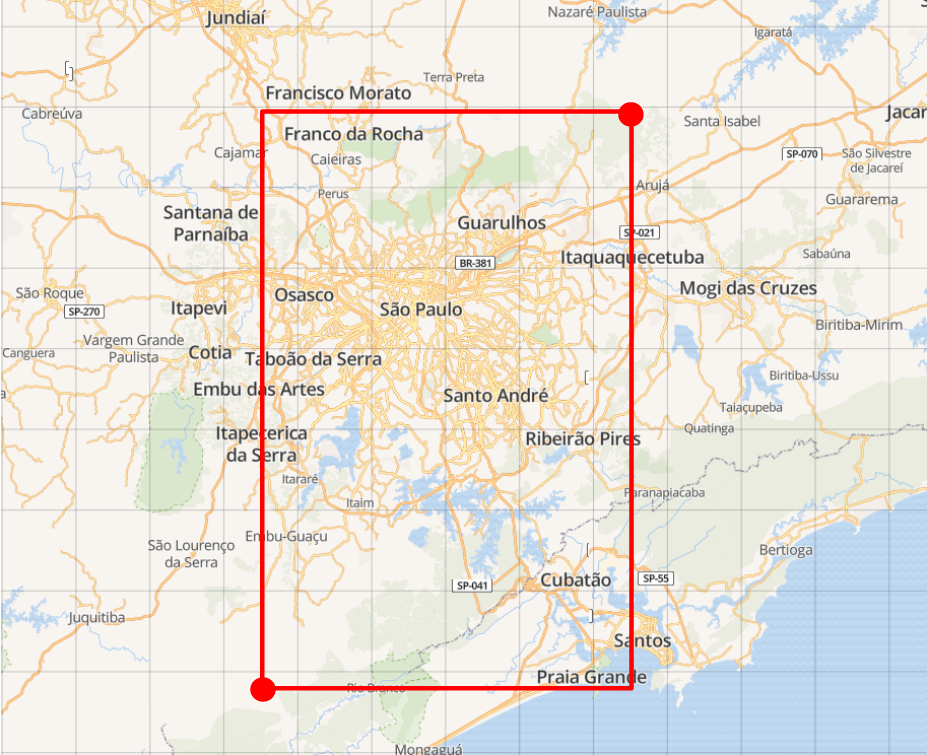}
\caption{S\~ao Paulo}
\label{fig:bb_sp}
\end{figure}

The selected target scenarios to this study are the two most populous and active Brazilian cities in Twitter, Rio de Janeiro and São Paulo, and the cities corresponding bounding-boxes used in the collecting process are presented in Figures~\ref{fig:bb_rio} and~\ref{fig:bb_sp}.

Messages sum up a total of 12.5M and 6.3M tweets for Rio de Janeiro and São Paulo, respectively, however during the retrieving process, it was identified that not all collected tweets were located inside the defined bounding-boxes. Indeed, according to the Twitter Streaming API documentation\footnote{\url{https://dev.twitter.com/streaming/overview/request-parameters#locations}}, a tweet is retrieved having into consideration two different heuristics: (1) If the coordinates field is populated, the values there will be tested against the bounding-box; (2) If the coordinates field is empty but place is populated, the region defined in place is checked for intersections against the locations bounding-box. Any overlapping will yield a positive match.

If a user tags his precise location within the posted tweet, then the first heuristic is applied. Otherwise, Twitter provides a list of places in which a user can select one of them and tags the tweet with it. Each \texttt{place} is composed by a bounding-box, which is used in the second heuristic of the API. In this heuristic, a matching between the collecting filter's and tweet's bounding-boxes is made, and if there is any overlap, a positive match is yielded and the tweet is retrieved. For example, if a tweet that has a \texttt{place} such as Brazil and the collecting filter bounding-box is defined for SP, all tweets from place Brazil will be retrieved to the final dataset, regardless the fact some tweets are posted very far away from the target bounding-box.

\begin{table*}[!htbp]
\small
\centering
\caption{Datasets composition}
\label{tab:datasets}
\begin{tabular}{l|S[table-format=8.0]|S[table-format=8.0]|S[table-format=7.0]|c|c|c}
\hline
\textbf{City}  & \textbf{All} & \textbf{PT} & \textbf{Non-PT} & \textbf{\begin{tabular}[c]{@{}c@{}}In \\ Bounding-Box\end{tabular}} & \textbf{\begin{tabular}[c]{@{}c@{}}Out\\ Bounding-Box\end{tabular}} & \textbf{\begin{tabular}[c]{@{}c@{}}PT and \\ In Bounding-Box\end{tabular}} \\ \hline
Rio de Janeiro & 12,531,000 & 10,570,000 & 1,961,000 & 8,644,000 & 3,886,000 & 7,353,000 \\ \hline
S\~ao Paulo & 6,352,000 & 4,886,000 & 1,466,000 & 4,247,000 & 2,105,000 & 3,313,000 \\ \hline
\end{tabular}
\end{table*}

Since this study is conducted under real-world scenarios, the reliability of data is a mandatory characteristic and so, a solution for the matching problem was found. Using the default Twitter bounding-boxes for places RJ and SP, we calculated which bounding-boxes were entirely inside the cities. The final composition of the datasets is presented in Table~\ref{tab:datasets}, and the results of the filtering process shown that almost 6M tweets were not located inside the bounding-boxes of the cities. The partial datasets composed by Portuguese tweets and located inside the cities bounding-boxes were used to conduct the proposed experiments in this study. Summing up, datasets present a total of 7.3M and 3.3M for Rio de Janeiro and São Paulo, respectively.

\subsection{Exploratory analysis}


Due to the volume of the datasets, some analytics were performed in order to gain better knowledge with respect to its intrinsic characteristics. A tweet provides some fields of interest such as text message, date of creation, language, and the \emph{entities}, which are constantly analysed in several data analytics systems. An \emph{entity} is metadata and additional contextual information contained in the tweet and is composed by the \emph{hashtags}, \emph{user mentions}, \emph{urls} and \emph{media} fields. For both scenarios, RJ and SP, we count a number of tweets containing this kind of information. Table~\ref{tab:entities} present results for the \emph{entities} count and it is possible to verify that \emph{urls} and \emph{user mentions} are most used ones over both datasets. Such information shows that users tend to tag another ones in a message meaning that tweets are used as a mean of communication, which is the basis of a microblogging platform.

\begin{table*}[!htbp]
\small
\centering
\caption{Datasets entities statistics}
\label{tab:entities}
\begin{tabular}{l|c|S[table-format=2.0]|c|c|S[table-format=7.0]|c|c|c}
\hline
\multirow{2}{*}{\textbf{City}} & \multicolumn{2}{c|}{\textbf{Hashtags (\#)}} & \multicolumn{2}{c|}{\textbf{User Mentions (@)}} & \multicolumn{2}{c|}{\textbf{URLs}} & \multicolumn{2}{c}{\textbf{Media}} \\ \cline{2-9} 
 & \textbf{Total} & \textbf{\%} & \textbf{Total} & \textbf{\%} & \textbf{Total} & \textbf{\%} & \textbf{Total} & \textbf{\%} \\ \hline
\textbf{Rio de Janeiro} & 525,550 & 5 & 1,340,334 & 13 & 1,509,742 & 14 & 389,864 & 4 \\
\textbf{São Paulo} & 585,365 & 12 & 1,072,566 & 22 & 885,369 & 18 & 302,579 & 6 \\ \hline
\end{tabular}
\end{table*}

\begin{figure}[!htbp]
\centering
\includegraphics[width=0.8\linewidth]{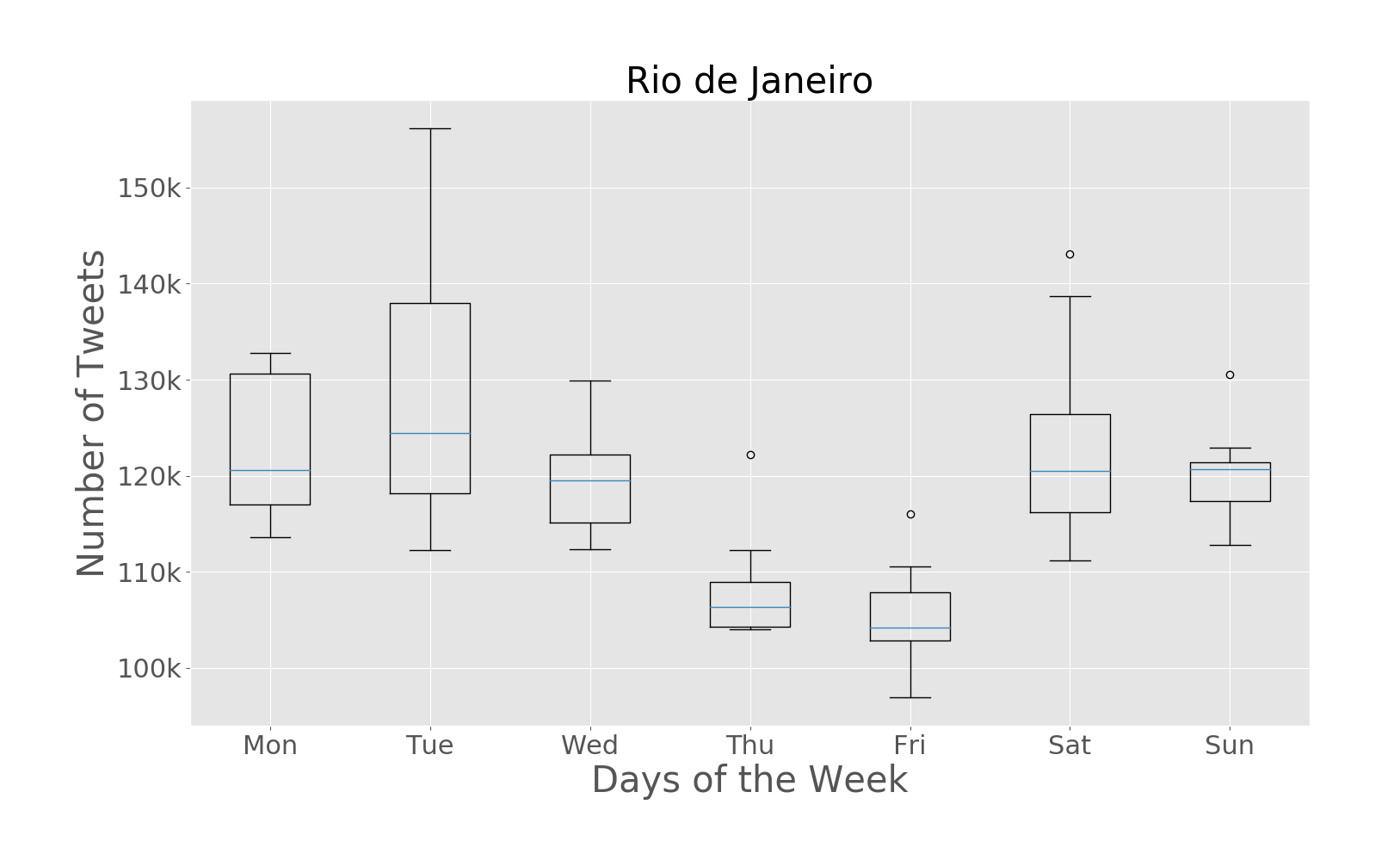}
\includegraphics[width=0.8\linewidth]{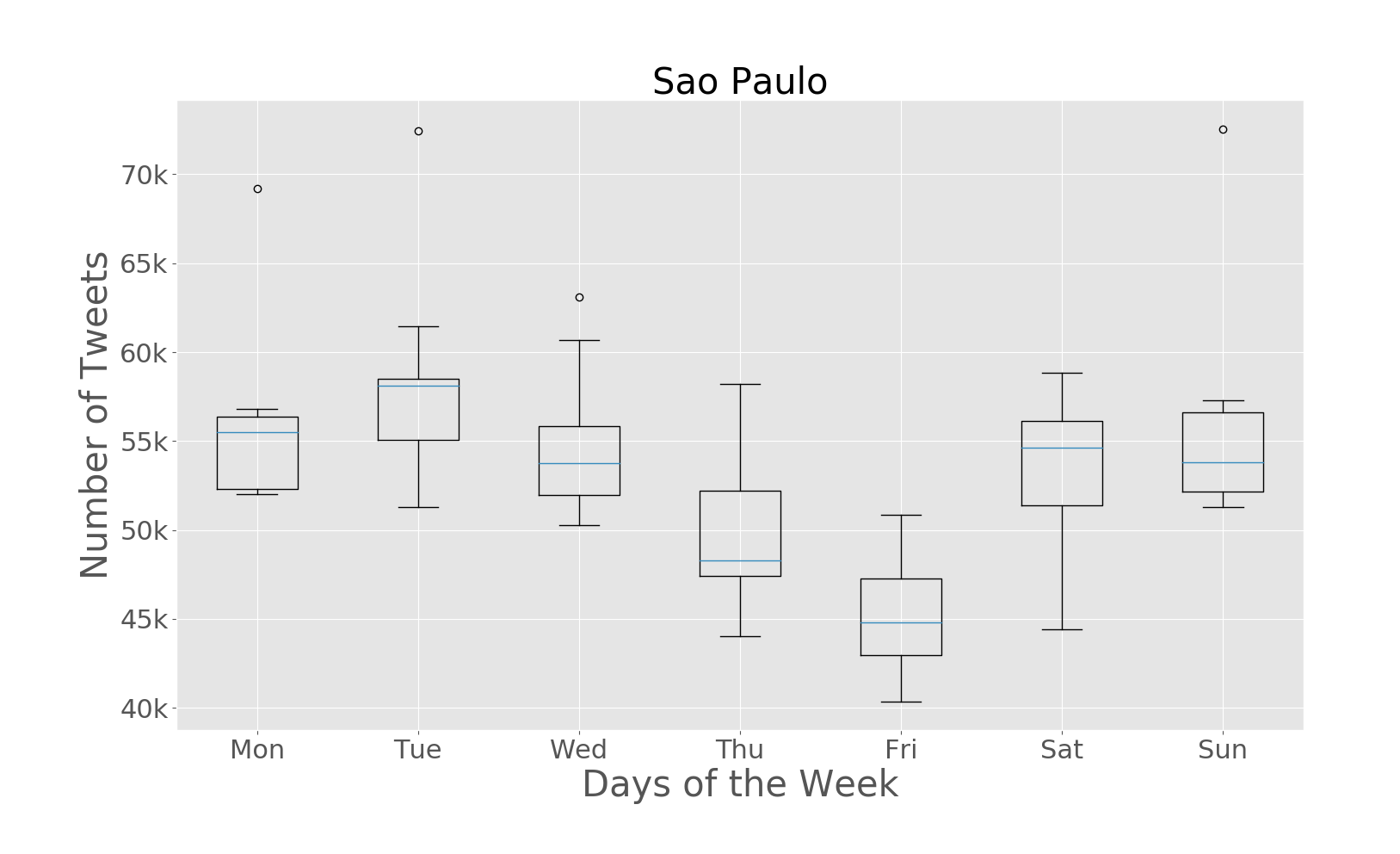}
\caption{Days of the week box-plots to the selected portion of the datasets}
\label{fig:box_plot_daily}
\end{figure}

The grouping of tweets by the day of the week is illustrated in Figure~\ref{fig:box_plot_daily} and particular points can be observed and considered uncommon. For both cities distributions and with respect to geo-located tweets, Friday is the less active day while the major activity occurs in the first days of the week. This is a strange phenomenon since Friday is the transition of the labour week days to the weekend and people could use more the microblog service to share their free time.

Regarding the hours of the day, usually there is a continuous growth between 16h and 23h. As expected, during the night the activity in Twitter is very limited for both cities.

\begin{figure}
\centering
\includegraphics[width=0.8\linewidth]{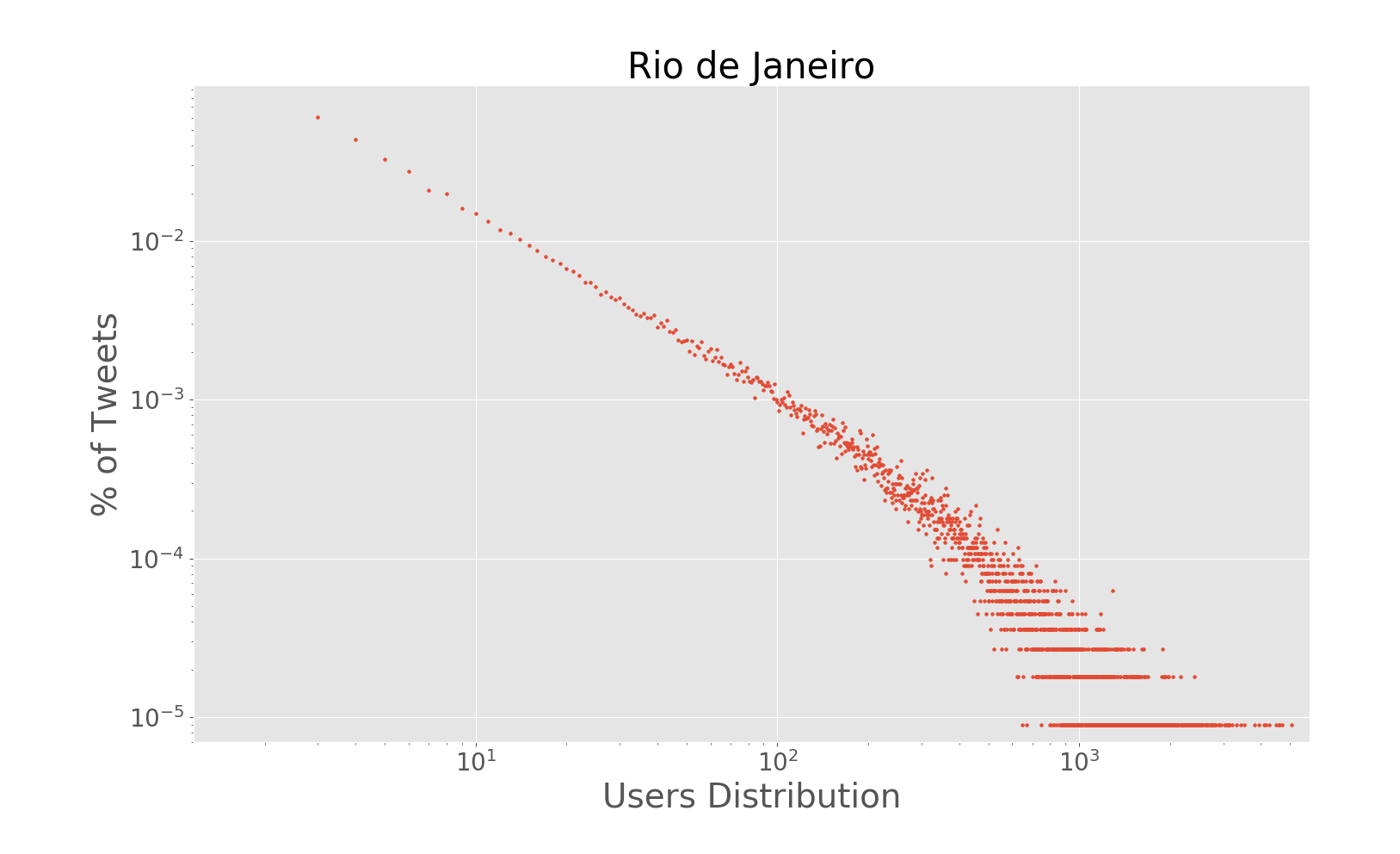}
\includegraphics[width=0.8\linewidth]{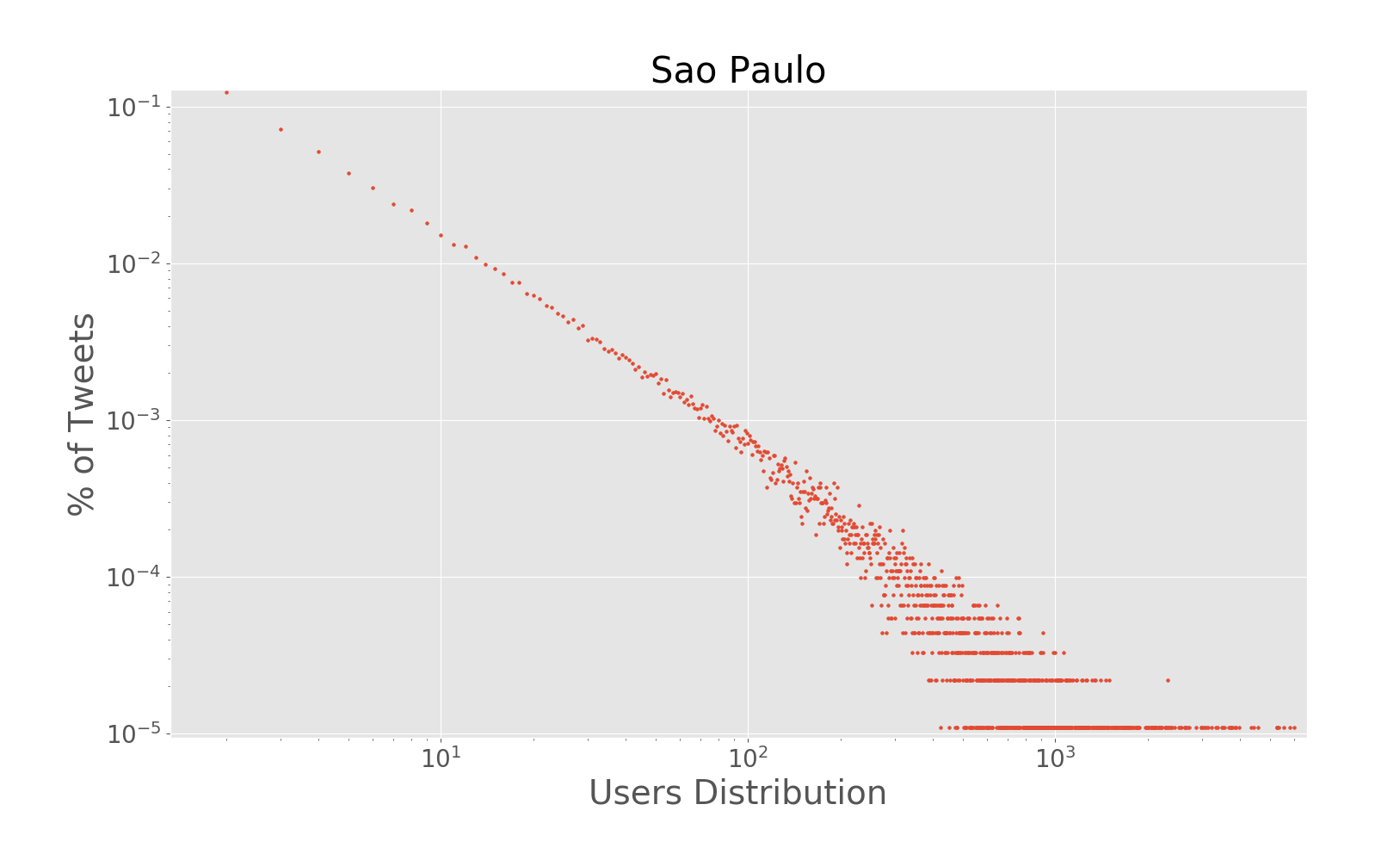}
\caption{Log-log distribution of Twitter users per tweets posted}
\label{fig:users_distribution}
\end{figure}

Another statistic calculated was the number of distinct users and the distribution of number of tweets per user in the whole datasets. In total, the RJ dataset contained 111,130 distinct users while the SP dataset had only 90,908 individuals. Such numbers are not indicators that Rio de Janeiro is a more active city than S\~ao Paulo since the bounding-boxes areas are not equal. Figure~\ref{fig:users_distribution} shows a similar behaviour regarding the distribution of the users per number of tweets. In both scenarios, it is possible to establish a correlation between a power law distribution and the frequency of tweets per user. The notable long tail shows us the existence of a large number of occurrences of users whose activity on Twitter is much lower.

\section{Topic modeling}
The main goal of this study is the characterization of tweets in two different Brazilian cities. We used an unsupervised learning approach to tackle the task of topic modelling in order to compare both cities and see if there are differences between subjects people talked about. Automatic characterization of text messages is a laborious and time-consuming task since it is necessary to assure the right level of abstraction in the learning model; very much similar to human minds, which essentially present a bounded rationality nature, our learning model needs to be trained in order to assimilate the necessary knowledge and perform the appropriate analogies so as to discover different topics within the tweets' contents. The premises to implement such a mechanism are presented and discussed in the following subsections.

\subsection{Latent Dirichlet Allocation (LDA)}
Blei et al.~\cite{blei2003latent} have developed a generative statistical model that makes possible the discovery of unknown groups and its similarities over any dataset. The model tries to identify what topics are present in a document by observing all the words that compose it, producing as the final result a topic distribution. An interesting point in this model is that the only features it analyses are the words passing through the training process. The model takes into consideration two different distributions - distribution of words over topics and distribution of topics over the documents - being each document seen as a random mixture of topics and each topic as a distribution of words.

\subsection{Data Preparation}
Each tweet of both datasets was submitted to a required group of pre-processing operations in order to train the LDA model and proceed with the experiments. The pre-processing steps were the ones detailed below.

\begin{itemize}
\item \textbf{Lowercasing:} Every message presented in a tweet was converted into lowercase;
\item \textbf{Cleaning Entities and Numbers:} Removing URLs, user mentions, hashtags and digits from the text message;
\item \textbf{Lemmatization:} Transformation of plural words into singular ones;
\item \textbf{Transforming repeated characters:} Sequences of characters repeated more than three times were transformed, e.g. "loooool" was converted to "loool";
\item \textbf{Punctuation Removal:} Every punctuation was removed as well as smiles or even \emph{emojis};
\item \textbf{Stop Words Removal:} The removing of this kind of words was made using the Portuguese NLTK dictionary;
\item \textbf{Short Tokens Removal:} Words such as 'kkk', 'aaa', 'aff' and other of the same style were removed.
\end{itemize}

After the data preparation phase, 772,017 tweets have their message empty which concludes that its content was irrelevant for the final experiment phase.

\section{EXPERIMENTAL SETUP}
\begin{figure}
\centering
\includegraphics[width=0.7\linewidth]{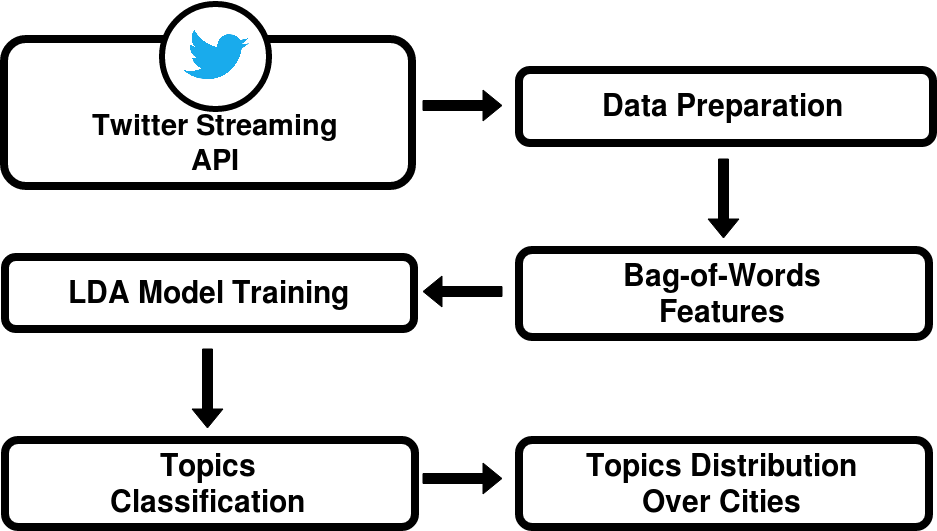}
\caption{Framework pipeline of operations}
\label{fig:pipeline}
\end{figure}

In this section, we describe the strategy chosen to create the text representation features for each document as well as the parametrization set in our model to conduct the study. The whole pipeline operations can be observed in Figure~\ref{fig:pipeline}, where it is represented the flow of the data from its collection method to the final distribution of topics over the tweets in order to characterize each of it.

\subsection{Features - Bag-of-words}
Topic modelling requires, like other learning models, a group of features to be trained. In this case, we used bag-of-words representation matrix - which represents each document as a frequency vector according to the number of occurrences of each word in the message. The set of features is limited to a dictionary containing 10,000 words and it only took in consideration uni-grams in the message content and words that occur in a maximum percentage of 40$\%$ in the whole dataset, avoiding common words that are not removed in the filtering process using NLTK Stop Words list. The minimal occurrence value for a word being considered was set to 10. 

\subsection{LDA model parametrization}
In order to understand and see the LDA model performance, we set five different numbers for the topics results parameter of the training process: 5, 10, 20, 25 and 50 topics, being this the one with better results. The number of iterations to train the model was set to 20. Each tweet in the datasets was treated as a single document comprehending that, in total, 6,580,983 different documents were used in the model training process.

\section{Results and Discussion}
\label{sec:results}

To evaluate the experimental results obtained for each model, a list of the most frequent 50 words for each topic was extracted. We also selected and manually analysed a random sample of 200 tweets for each topic. This sampling was done in order to get better consistency and trustiness about the classification and characterization of the tweets.

The model trained to get 50 topics was the one with the largest number of distinct topics between them, however, there were topics which theme was the same (e.g. Love and Romance Problems or Brazilian Football \textit{versus} European Football). Within this, such groups were aggregate into the same topic, \textit{Relationships} and \textit{Sports and Games}, respectively. After this grouping process, a total of 29 different topics was achieved. 

Some tweets that have added complexity to our classification objective, such as, for example, "\textit{queria namorar um mano parecido com o josh}" (Relationship) and "\textit{como eu queria meus amigos aqui agora cmg}" (Friendship), raised some doubts about which topic this tweets may belong: Relationship, Friendship or even Actions or Intentions. In a perspective of context, the first tweet belongs to the theme \texttt{flirt}, which is directly related to Relationship. The theme on the second tweet is missing the company of friends, i.e. conviviality, which is related to Friendship. The decision of joining the two topics was due to the proximity between them, which have as content both types of tweets talking about love/relationship and friendship, and, having this considered, both topics should be aggregated in order to assure the desired coherency in the classification.

The final set of topics (50 topics) to be considered was selected accordantly to the most recurring subjects. The final classification and details associated with the whole dataset for each city are presented in Table~\ref{tab:final_classification}. Almost every topic demonstrated a balanced distribution, with exception of \textit{Relationships and Friendship} and \textit{Personal Feelings} for Rio de Janeiro and S\~ao Paulo, respectively. The difference that appears in these topics is a consequence of the final grouping process since there was a considerable number of words been shared among them. This issue complicated our classification task, compelling to a high amount of undesired aggregations.

\begin{table*}[!htbp]
\centering
\caption{Final results of the aggregation of the LDA topics}
\label{tab:final_classification}
\resizebox{0.7\textwidth}{!}{\begin{tabular}{l|S[table-format=7.0]S[table-format=2.2]|S[table-format=7.0]S[table-format=2.2]|rS[table-format=2.2]}
\hline
\multicolumn{1}{c|}{\multirow{2}{*}{\textbf{Topic Group}}} & \multicolumn{2}{c|}{\textbf{Rio de Janeiro}} & \multicolumn{2}{c|}{\textbf{S\~ao Paulo}} & \multicolumn{1}{c}{\multirow{2}{*}{\textbf{Diff (\%)}}} \\ \cline{2-5}
\multicolumn{1}{c|}{} & \textbf{No. Tweets} & \textbf{Percentage (\%)} & \textbf{No. Tweets} & \textbf{Percentage (\%)} & \multicolumn{1}{c}{} \\ \hline
Academic Activities & 101,590 & 1.54 & 90,616 & 3.30 & -1.76 \\
Actions or Intentions & 600,030 & 9.12 & 128,710 & 4.69 & \textbf{+4.43} \\
Antecipation and Socialising & 132,606 & 2.01 & 0 & 0.00 & \textbf{+2.01} \\
BBB17 & 122,054 & 1.85 & 68,385 & 2.49 & -0.64 \\
Body, Appearances and Clothes & 160,342 & 2.44 & 71,447 & 2.60 & -0.17 \\
Food and Drink & 167,204 & 2.54 & 58,407 & 2.13 & +0.41 \\
Health & 119,013 & 1.81 & 0 & 0.00 & \textbf{+1.81} \\
Holidays and Weekends & 104,695 & 1.59 & 79,610 & 2.90 & -1.31 \\
Informal Conversations & 272,502 & 4.14 & 138,848 & 5.06 & -0.92 \\
Live Shows, Social Events and Nightlife & 359,342 & 5.46 & 140,240 & 5.11 & +0.35 \\
Mood & 139,287 & 2.12 & 138,399 & 5.04 & \textbf{-2.92} \\
Movies and TV & 285,198 & 4.33 & 39,778 & 1.45 & \textbf{+2.89} \\
Music and Artists & 84,407 & 1.28 & 78,142 & 2.85 & 1.56 \\
Negativism, Pessimism and Anger & 229,104 & 3.48 & 183,050 & 6.67 & \textbf{-3.18} \\
Numbers, Quantities and Classification & 86,897 & 1.32 & 78,160 & 2.85 & -1.53 \\
Optimism and Positivism & 106,714 & 1.62 & 39,725 & 1.45 & +0.18 \\
Personal Fellings & 375,735 & 5.71 & 532,331 & 19.38 & \textbf{-13.67} \\
Politics & 81,254 & 1.23 & 46,758 & 1.70 & 0.47 \\
Relationships and Friendship & 1,524,804 & 23.17 & 187,541 & 6.83 & \textbf{+16.34} \\
Religion & 183,174 & 2.78 & 66,788 & 2.43 & +0.35 \\
Routine Activities & 334,216 & 5.08 & 82,421 & 3.00 & +2.08 \\
Slang and Profinities & 241,676 & 3.67 & 44,620 & 1.62 & +2.05 \\
Social Media Applications & 105,809 & 1.61 & 44,073 & 1.60 & +0.01 \\
Sport and Games & 382,479 & 5.81 & 133,047 & 4.84 & +0.97 \\
Tourism and Places & 59,288 & 0.90 & 86,519 & 3.15 & -2.25 \\
Transportation and Travel & 130,261 & 1.98 & 63,923 & 2.33 & -0.35 \\
Weather & 91,302 & 1.39 & 42,588 & 1.55 & -0.16 \\
Shopping & 0 & 0.00 & 44,470 & 1.62 & \textbf{-1.62} \\
Voting & 0 & 0.00 & 37,687 & 1.37 & \textbf{-1.37} \\ \hline
\end{tabular}}
\end{table*}

Additionally to the manual verification of a sample of tweets for each topic, we also produced a temporal week day distribution,  with the objective to observe if some topics had more mentions in certain days than others.

For making such observations some assumptions were made in relation with some \textit{hot} topics. More specifically, we think that is valid to assume that people will talk more about \textit{Religion} in the weekend, since they go to the church in those days. The same result is likely to happen for topics like \textit{Holidays and Weekends} or \textit{Sports and Games}, since events related to this thematics occur during specific time-frames. 

\begin{figure}[!tbp]
	\centering
	\subfloat[Rio de Janeiro]{\includegraphics[width=2.6in]{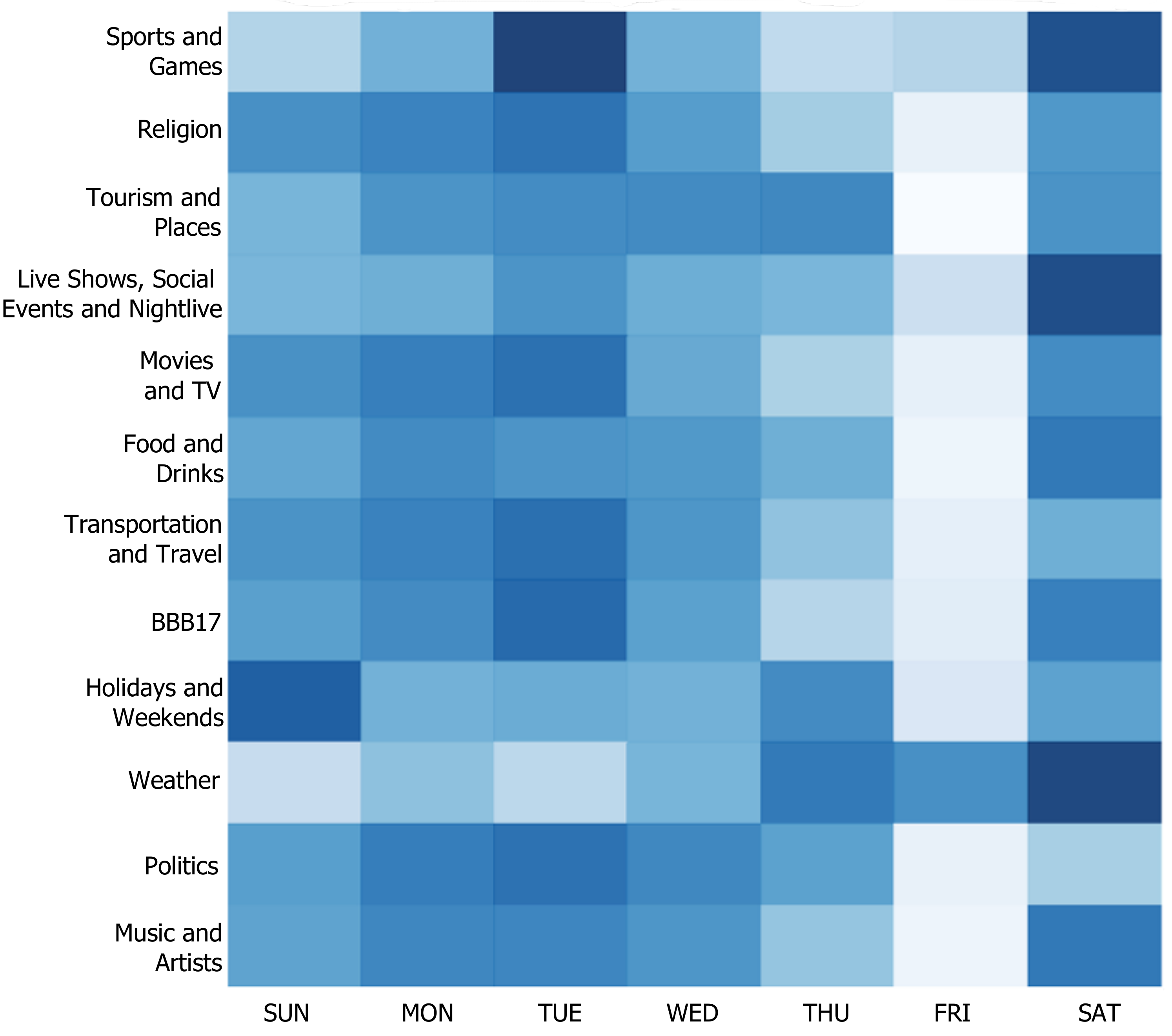}%
		\label{subfig:sp}}
	\hfil
	\subfloat[São Paulo]{\includegraphics[width=2.6in]{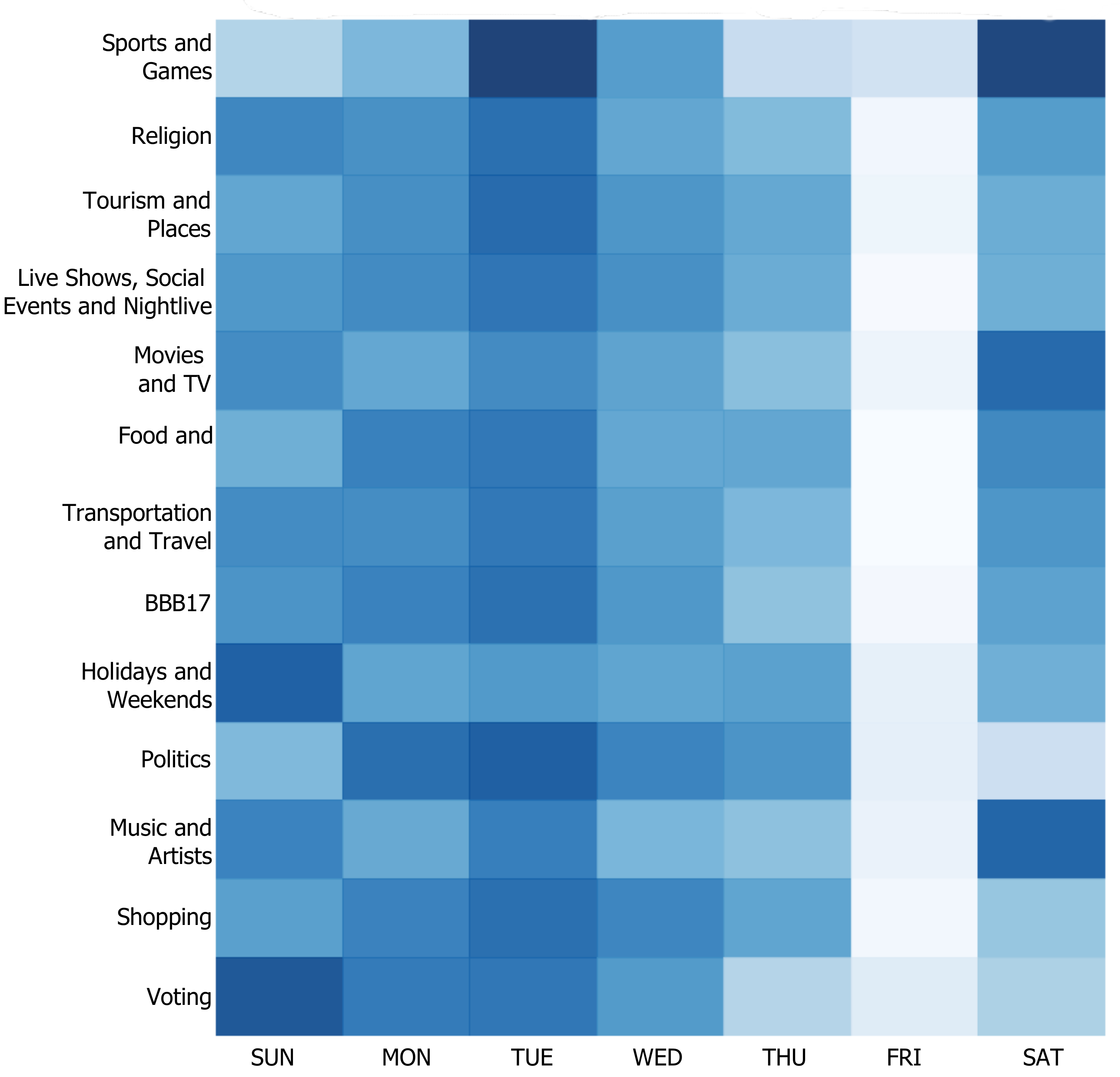}%
		\label{fig_second_case}}
	\caption{Day-of-the-week activity per each topic in both cities}
	\label{fig:heatmap_distribution}
\end{figure}

Only 12 topics of the finals 29 were selected for this part of the study, predicting them and comparing the final results, such as, but not limited to, \textit{Sports and Games}, \textit{Religion}, \textit{Holidays and Weekends}, \textit{Movies and TV}, \textit{Live Shows, Social Events and Nightlife}. The temporal distribution is shown in Figure~\ref{fig:heatmap_distribution} as a heat map, where each row is independent of the others.

The necessity of applying such restrictions is due to the need of seeing in which days each topic is more talked about. For both cities the topic \textit{Sports and Games} is more mentioned in Tuesdays and Saturdays. Indeed, this observation correlates with the days that topic-related events happen. Namely, Tuesdays and Wednesday correspond to the days when the \textit{UEFA Champions League} competition happens and Saturdays and Sundays to the days of \textit{Brazilian Football League} games. \textit{Holidays and Weekends} was a topic with interesting results regarding the temporal distribution, presenting Sundays as the day where more people talk about it. 

Nonetheless, it is worth noting that our model had successfully discovered a topic related to \textit{Big Brother Brazil 2017} (BBB17), a well-known reality show. The amount of geo-located tweets concerning this topic was considerable (1.85\% and 2.49\%, in RJ and SP, respectively), raising the question about what led people to geo-located them in such topic.

\section{Conclusions}

This research demonstrates the capability of our framework to handle different types of analysis under unregulated and non-conventional data such as the content found in most social media. The application of topic modelling technique to tweets from two different cities enables interesting comparisons between them since the whole analytics process accounts for what inhabitants talk about in their social networks. Through these analyses, cities' services are capable of monitoring human behaviour, activity patterns as well as of identifying regions where there may be some levels of intolerance on certain topics, making it possible to trigger preventive measures to solve problems in those specific areas.

LDA models usually require documents of large size, or at least ones with higher complexity than a single tweet so as to yield appropriate performance. A traditional approach was followed considering each tweet as a document instead of trying to aggregate tweets in more complex documents taking into consideration some criteria, e.g. grouping messages by date and hour. All topics resulting from our approach are similar in both cities but two, which are unique for each of the selected scenarios. The percentage difference between similar topics was within the interval 0.16-4.43\% evidencing the fact that both cities are also similar besides different factors that characterize each other: population, culture, lifestyle and also the region where the city is located in.

In spite of the analysis carried out and reported in this paper, we can not assure that inside a topic other encapsulated topics might exist. The resulting amount of tweets for each topic was extremely high, turning a one-by-one verification into a very laborious and time-consuming process. Therefore, our classification approach was limited to the verification of the top-50 words and the manual identification of such words in samples of 200 tweets per topic. Such a limitation, nonetheless, did not prevent us to draw important conclusions relative to the results we have obtained after the application of our proposed solutions, as discussed in the Section \ref{sec:results}.

Future direction for this research will include an application of spatio-temporal aggregation methods over both datasets in order to create more complex documents and verify whether results can be different taking into consideration temporal and spatial factors. To pursue this, it is required that a large dataset for both cities is available, which is expectable only in mid- to long-term. A possible future evaluation approach to be considered is the mapping of topics over specific areas of the cities, such as the identification of topics related to beaches alongside the coastal area in Rio de Janeiro, or the identification of transportation and travel topics over the metropolitan area in São Paulo. Finally, it is also necessary to explore other classification/evaluation approaches to enhance robustness, consistency, and efficiency of the topic modelling routine; one possible solution is the method aforementioned in Section~\ref{related_work}, which considers the addition of an extra layer to endow the framework with supervised labelling capabilities.

\section*{Acknowledgements}
This work was partially supported by Project "TEC4Growth - Pervasive Intelligence, Enhancers and Proofs of Concept with Industrial Impact/NORTE-01-0145-FEDER-000020".

\bibliographystyle{unsrt}
\bibliography{refs}

\end{document}